\documentstyle[preprint,aps]{revtex}
\begin{document}

\draft

\title{Mean magnetic field renormalization and Kolmogorov's energy spectrum in MHD
turbulence}
\author{Mahendra\ K.\ Verma \thanks{email: mkv@iitk.ernet.in}}

\address{Department of Physics, Indian Institute of Technology,
Kanpur  --  208016, INDIA}

\date{15 March, 1998}
\maketitle

\begin{abstract}
In this paper we construct a self-consistent renormalization group procedure
for MHD\ turbulence in which small wavenumber modes are averaged out, and
effective mean magnetic field at large wavenumbers is obtained. In this
scheme the mean magnetic field scales as $k^{-1/3}$ , while the energy
spectrum scales as $k^{-5/3}$ similar to that in fluid turbulence. We also
deduce from the formalism that the magnitude of cascade rate decreases as
the strength of the mean magnetic field is increased.
\end{abstract}

\vspace{1.5cm}


\section{Introduction\ }

Kolmogorov hypothesized that the energy spectrum $E(k)$ of fluid turbulence
in the inertial range is isotropic and is a power law with a spectral index
of $-5/3$, i.e., 
\begin{equation}
\label{Kolmfl}E(k)=K_{Ko}\Pi ^{2/3}k^{-5/3} 
\end{equation}
where $K_{Ko}$ is an universal constant called Kolmogorov's constant, $k$ is
the wavenumber, and $\Pi $ is the nonlinear energy cascade rate. Note that $%
\Pi $ is equal to the dissipation rate and also the energy supply rate of
the fluid. Experiments \cite{Gran}, simulations \cite{FrisSul}, and some of
the analytical calculations based on Direct interaction approximation \cite
{Krai59,Lesl}, renormalization group (RG) techniques \cite
{FNS,YakhOrs,Roni,Mac,MacSha,MacWat,ZhouVah}, self-consistent mode coupling 
\cite{Bhat91} etc. are in good agreement with the above phenomenology.

In this paper we will discuss the energy spectrum in magnetohydrodynamic
(MHD) turbulence. In MHD there are two fields, the velocity field {\bf u}
and the magnetic field ${\bf B=B}_0+{\bf b}$, where ${\bf B}_0$ is the mean
magnetic field or the magnetic field of the large eddies, and {\bf b }is the
magnetic field fluctuation. One usually uses Els\"asser variables ${\bf z}%
^{\pm }={\bf u\pm b}$. Here the magnetic field has been written in velocity
units ($b/\sqrt{4\pi \rho }$, where $\rho $ is the density of the fluid). We
also assume that the plasma is incompressible.

There are two time-scales in magnetofluid: (i) nonlinear time-scale $%
1/(kz_k^{\pm })$ (similar to that in fluid turbulence) and (ii) Alfv\'en
time-scale $1/(kB_0)$. Kraichnan \cite{Krai65} and Dobrowolny et al. \cite
{Dobr80} argued that the interacting $z_k^{+}$ and $z_k^{-}$ modes will get
separated in one Alfv\'en time-scale because of the mean magnetic field.
Therefore, they chose Alfv\'en time scale $\tau _A=(kB_0)^{-1}$ as the
relevant time-scale and found that 
\begin{equation}
\label{Dobro}\Pi ^{+}\approx \Pi ^{-}\approx \frac 1{B_0}%
E^{+}(k)E^{-}(k)k^3=\Pi . 
\end{equation}
where $\Pi ^{\pm }$ \thinspace are the cascade rates of $z_k^{\pm }$. If $%
E^{+}(k)\approx E^{-}(k),$ then the above equation implies that 
\begin{equation}
E^{+}(k)\approx E^{-}(k)\approx \left( B_0\Pi \right) ^{1/2}k^{-3/2} 
\end{equation}
In absence of mean magnetic field, the magnetic field of the largest eddy
was taken as $B_0$. Kraichnan \cite{Krai65} also argued that the fluid and
magnetic energies are equipartitioned. The above phenomenology is referred
to as Dobrowolny et al.'s generalized Kraichnan (KD) phenomenology.

If the nonlinear time-scale $\tau _{NL}^{\pm }\approx kz_k^{\mp }$ is chosen
as the interaction time-scales for the eddies $z_k^{\pm }$, we obtain 
\begin{equation}
\Pi ^{\pm }\approx \left( z_k^{\pm }\right) ^2\left( z_k^{\mp }\right) k, 
\end{equation}
which in turn leads to 
\begin{equation}
\label{eq:mhd}E^{\pm }(k)=K^{\pm }(\Pi ^{\pm })^{4/3}(\Pi ^{\mp
})^{-2/3}k^{-5/3}, 
\end{equation}
where $K^{\pm }$ are constants, which we will refer to as Kolmogorov's
constants for MHD turbulence. Because of its similarity with Kolmogorov's
fluid turbulence phenomenology, this phenomenology is referred to as
Kolmogorov-like MHD turbulence phenomenology. This phenomenology was first
given by Marsch \cite{Mars}, Matthaeus and Zhou \cite{MattZho}, and Zhou and
Matthaeus \cite{ZhouMat} (it is a limiting case of a more generalized
phenomenology constructed by Matthaeus and Zhou \cite{MattZho}, and Zhou and
Matthaeus \cite{ZhouMat}). It is implicit in these phenomenological
arguments that KD phenomenology is expected to hold when $B_0\gg \sqrt{%
kE^{\pm }(k)}$, while Kolmogorov-like phenomenology is expected to be
applicable when $B_0\ll \sqrt{kE^{\pm }(k)}$.

In the solar wind, which is a good testing ground for MHD turbulence
theories, Matthaeus and Goldstein \cite{MattGol} found that the exponent of
the total energy is $1.69\pm 0.08$, whereas the exponent of the magnetic
energy is $1.73\pm 0.08$, somewhat closer to 5/3 than 3/2. This is more
surprising because $B_0\gg \sqrt{kE^{\pm }(k)}$ for inertial range
wavenumbers in the solar wind. The numerical simulations also tend to
indicate that the Kolmogorov-like phenomenology, rather than KD
phenomenology, is probably applicable in MHD turbulence \cite{Verm96sim}.
Hence, the comparison of the solar wind observations and simulation results
with the phenomenological predictions appears to show that there are some
inconsistencies in the phenomenological arguments given above. To resolve
these inconsistencies, we have attempted to examine the MHD equations using
renormalization group analysis.

For fluid turbulence Forster et al. \cite{FNS} and Yakhot and Orszag \cite
{YakhOrs} have applied dynamical RG procedure in which a forcing term with a
power law distribution in wavenumber space is introduced. McComb \cite{Mac},
McComb and Shanmugasundaram \cite{MacSha}, McComb and Watt \cite{MacWat},
and Zhou et al. \cite{ZhouVah} applied a self-consistent RG procedure that
yields Kolmogorov's energy spectrum. For MHD turbulence, Fournier et al. 
\cite{FourSul} and Camargo and Tasso \cite{Cama} have used RG procedure
similar to that of Forster et al. \cite{FNS} and Yakhot and Orszag \cite
{YakhOrs}. In all these schemes the averaging is done over the small scales
(based on Wilson's approach in his Fourier space RG). Till date the RG
methods applied to MHD turbulence do not find direct evidence of
Kolmogorov-like power law in MHD\ turbulence. In a more recent work, Verma
and Bhattacharjee \cite{VermBha} have applied Kraichnan's DIA \cite
{Krai59,Lesl} to MHD turbulence and obtained the Kolmogorov's constant for
MHD, but in Verma and Bhattacharjee's work $k^{-5/3}$ energy spectra was
assumed, and an artificial cutoff was introduced for the self energy
integral.

In this paper we construct a self-consistent RG procedure similar to that
used by McComb \cite{Mac}, McComb and Shanmugsundaram \cite{MacSha}, McComb
and Watt \cite{MacWat}, and Zhou et al. \cite{ZhouVah} for fluid turbulence.
However, one major difference is that we integrate the small wavenumber
modes instead of large wavenumber mode integration used by earlier authors.
In our procedure we obtain the effective mean magnetic field $B_0(k)$ as we
go from small wavenumbers to large wavenumbers. At small wavenumbers the MHD
equations are approximately linear. During the RG process, the effects of
the nonlinear terms in the small wavenumber shells is translated to the
modification of $B_0(k)$ at larger wavenumbers.

We postulate that the effective mean magnetic field is the magnetic field of
the next-largest eddy contrary to the KD phenomenology where the effective
mean magnetic at any scale is constant. To illustrate, for Alfv\'en waves of
wavenumber $k$, the effective magnetic field $B_i(k)$ (after $i$th iteration
of the RG procedure defined below) will be the magnetic field of the eddy of
size $k/10$ or so. This argument is based on the physical intuition that for
the scattering of the Alfv\'en waves at a wavenumber $k$, the effects of the
magnetic field of the next-largest eddy is much more than that of the
external field. The mean magnetic field at the largest scale will simply
convect the waves; the local inhomogeneities contribute to the scattering of
waves which leads to turbulence (note that in WKB method, the local
inhomogeneity of the medium determines the amplitude and the phase
evolution). In our self consistent scheme we find that $B_0$ appearing in
the Kraichnan's or Dobrowolny et al.'s argument must be $k$ dependent. The
substitution of $k$ dependent $B_0(k)$ leads to $k^{-5/3}\,$energy spectra,
which is consistent with the solar wind observations and the simulation
results. We will describe these ideas in more detail in the following
section.

The normalized cross helicity $\sigma _c$, defined as $%
(E^{+}-E^{-})/(E^{+}+E^{-})$, and the Alfv\'en ratio $r_A$, defined as the
ratio of fluid energy and magnetic energy, are important factors in MHD
turbulence. For simplicity of the calculation, we have taken $%
E^{+}(k)=E^{-}(k)$ and $r_A=1$. These conditions are met at many places in
the solar wind and in other astrophysical plasmas.

\section{Calculation}

The MHD equation in the Fourier space is \cite{Krai65} 
\begin{equation}
\label{mhdeqn}\left( -i\omega \mp i\left( {\bf B}_0\cdot {\bf k}\right)
\right) z_i^{\pm }({\bf k,}\omega )=-iM_{ijm}({\bf k})\int d{\bf p}d\omega
^{\prime }z_j^{\mp }({\bf p},\omega ^{\prime })z_m^{\pm }({\bf k-p},\omega
-\omega ^{\prime }) 
\end{equation}
where 
\begin{equation}
M_{ijm}({\bf k})=k_jP_{im}({\bf k});\ P_{im}({\bf k})=\delta _{im}-\frac{%
k_ik_m}{k^2}, 
\end{equation}
Here we have ignored the viscous terms. The above equation will, in
principle, yield an anisotropic energy spectra (different spectra along and
perpendicular to ${\bf B}_0$). Solving anisotropic equations is quite
complicated. Therefore, we modify the above equation to the following form
to preserve isotropy: 
\begin{equation}
\label{mhdk}\left( -i\omega \mp i\left( B_0k\right) \right) z_i^{\pm }({\bf %
k,}\omega )=-iM_{ijm}({\bf k})\int d{\bf p}d\omega ^{\prime }z_j^{\mp }({\bf %
p,}\omega ^{\prime })z_m^{\pm }({\bf k-p,}\omega -\omega ^{\prime }) 
\end{equation}
This equation can be thought of as an effective MHD equation in an isotropic
random mean magnetic field.

In our RG procedure the wavenumber range $(k_0..k_N)$ is divided
logarithmically into $N$ shells. The $n$th shell is $(k_{n-1}..k_n)$ where $%
k_n=s^nk_o(s>1)$. In the following discussion, firstly we carry out the
elimination of the first shell $(k_0..k_1)$ and obtain the modified MHD\
equation. We then proceed iteratively to eliminate higher shells and get a
general expression for the modified MHD equation after elimination of $n$th
shell. The details of the renormalization group operation is as follows:

\subsection{RG Procedure}

\begin{enumerate}
\item  Decompose the modes into the modes to be eliminated $(k^{<})$ and the
modes to be retained $(k^{>}).$ In the first iteration $(k_0..k_1)=k^{<}$
and $(k_1..k_N)=k^{>}$. Note that $B_0(k)$ is the mean magnetic field before
the elimination of the first shell.

\item  We rewrite the Eq. (\ref{mhdk}) for $k^{<}$ and $k^{>}$. The equation
for $z_i^{\pm >}({\bf k},t)$ modes is 
\begin{equation}
\label{klarge}
\begin{array}{c}
\left( -i\omega \mp i\left( B_0k\right) \right) z_i^{\pm >}(
{\bf k},\omega )=-iM_{ijm}({\bf k})\int d{\bf p}d\omega ^{\prime }\left[
z_j^{\mp >}({\bf p},\omega ^{\prime })z_m^{\pm >}({\bf k-p},\omega -\omega
^{\prime })\right] + \\ \left[ z_j^{\mp >}(
{\bf p},\omega ^{\prime })z_m^{\pm <}({\bf k-p},\omega -\omega ^{\prime
})+z_j^{\mp <}({\bf p},\omega ^{\prime })z_m^{\pm >}({\bf k-p},\omega
-\omega ^{\prime })\right] + \\ \left[ z_j^{\mp <}({\bf p},\omega ^{\prime
})z_m^{\pm <}({\bf k-p},\omega -\omega ^{\prime })\right] 
\end{array}
\end{equation}
while the equation for $z_i^{\pm <}({\bf k},t)$ modes can be obtained by
interchanging $<$ and $>$ in the above equation.

\item  The terms given in the second and third brackets in the RHS of Eq. (%
\ref{klarge}) is calculated perturbatively. We perform ensemble average over
the first shell which is to be eliminated. We assume that $z_i^{\pm <}({\bf k%
},t)$ has a gaussian distribution with zero mean. Hence, 
\begin{equation}
\begin{array}{c}
\left\langle z_i^{\pm <}(
{\bf k},t)\right\rangle =0 \\ \left\langle z_i^{\pm >}({\bf k}%
,t)\right\rangle =z_i^{\pm >}({\bf k},\omega )
\end{array}
\end{equation}
and 
\begin{equation}
\left\langle z_s^{a<}({\bf p},\omega ^{\prime })z_m^{b<}({\bf q},\omega
^{\prime \prime })\right\rangle =P_{sm}({\bf p)}C^{ab}(p,\omega ^{\prime
})\delta {\bf (p+q)}\delta (\omega ^{\prime }+\omega ^{\prime \prime })
\end{equation}
where $a,b=\pm $. Also, the triple order correlations $\left\langle z_s^{\pm
<}({\bf k},\omega )z_m^{\pm <}({\bf p},\omega ^{\prime })z_t^{\pm <}({\bf q}%
,\omega ^{\prime \prime })\right\rangle $ are zero. We keep only the
nonvanishing terms to first order. For the relevant Feynmann diagrams, refer
to Zhou et al. \cite{ZhouVah}. Taking $r_A=1$ and $E^{+}(k)=E^{-}(k)$, the
Eq. (\ref{klarge}) becomes 
\begin{equation}
\label{klarge2}
\begin{array}{c}
\left( -i\omega \mp i\left( B_0k\right) \right) z_i^{\pm >}(
{\bf k},\omega )=-iM_{ijm}({\bf k})\int d{\bf p}d\omega ^{\prime }\left[
z_j^{\mp >}({\bf p},\omega ^{\prime })z_m^{\pm >}({\bf k-p},\omega -\omega
^{\prime })\right] + \\ \left( -i\right) ^2M_{ijm}(
{\bf k})\int_{{\bf p+q=k}}d{\bf q}d\omega ^{\prime }M_{mst}({\bf p)}P_{js}(%
{\bf q)}G^{\pm \pm }({\bf p},\omega ^{\prime })C^{\mp \mp <}({\bf q},\omega
-\omega ^{\prime })z_t^{\pm >}({\bf k},\omega )+ \\ \left( -i\right)
^2M_{ijm}({\bf k})\int_{{\bf p+q=k}}d{\bf q}d\omega ^{\prime }M_{mst}({\bf p)%
}P_{js}({\bf q)}G^{\pm \mp }({\bf p},\omega ^{\prime })C^{\mp \mp <}({\bf q}%
,\omega -\omega ^{\prime })z_t^{\pm >}({\bf k},\omega )
\end{array}
\end{equation}
where $G$ is the Green's function obtained from the equation 
\begin{equation}
G^{-1}(k,\omega )=\left( 
\begin{array}{cc}
-i\omega -ikB_0^{++}(k) & -ikB_0^{+-}(k) \\ 
ikB_0^{-+}(k) & -i\omega +ikB_0^{--}(k)
\end{array}
\right) .
\end{equation}

In deriving Eq. (\ref{klarge2}) we have neglected the contribution of the
triple nonlinearity $z_s^{\pm >}({\bf k},\omega )z_m^{\pm >}({\bf p},\omega
^{\prime })z_t^{\pm >}({\bf q},\omega ^{\prime \prime })$. McComb, McComb
and Shanmugsundaram, and McComb and Watt \cite{Mac,MacSha,MacWat} have also
ignored the triple nonlinearity for fluid turbulence.

\item  Since $r_A=1$ and $E^{+}(k)=E^{-}(k)$, we find that $%
B_0^{+-}(k)=B_0^{-+}(k)$. We also assume that the correlation functions $%
C^{\pm \pm }$ have the same frequency dependence as $G^{\pm \pm }$, i.e., 
\begin{equation}
C^{\pm \pm }(k,\omega ^{\pm })=\frac{C^{\pm \pm }(k)}{-i\omega ^{\pm }\mp
ikB_0^{\pm \pm }(k)}
\end{equation}
Note that $C^{\pm \pm }(k)=E^{\pm \pm }(k)/(4\pi k^2)$ in three dimensions.
From dynamical scaling arguments 
\begin{equation}
\omega ^{\pm }=\mp kB_0^{\pm \pm }(k)
\end{equation}
After some manipulations the Eq. (\ref{klarge2}) becomes $\,$ 
\begin{equation}
\begin{array}{c}
\left( -i\omega \mp i\left[ B_0(k)+\delta B_0^{\pm \pm }(k)\right] k\right)
z_i^{\pm >}(
{\bf k},t)\mp i\delta B_0^{\pm \mp }(k)z_i^{\mp >}({\bf k},t) \\ =M_{ijm}(%
{\bf k})\int d{\bf p}\left[ z_j^{\mp >}({\bf p},t)z_m^{\pm >}({\bf k-p}%
,t)\right] 
\end{array}
\end{equation}
where 
\begin{equation}
\label{pp}
\begin{array}{c}
\delta B_0^{\pm \pm }(k)=-k\int_{
{\bf p+q=k}}d{\bf q}\left( \frac{E(q)}{4\pi q^2}\right) \times  \\ \left[ 
\frac{a_2(k,p,q)\left( X_0^{\pm \pm }(p)+B_0^{\pm \pm }(p)\right)
-a_4(k,p,q)B_0^{+-}(p)}{2X_0(p)\left( kB_0^{\pm \pm }(k)+pX_0^{\pm \pm
}(p)-qX_0^{\pm \pm }(q)\right) }\right] 
\end{array}
\end{equation}
and 
\begin{equation}
\label{pm}
\begin{array}{c}
\delta B_0^{\pm \mp }(k)=-k\int_{
{\bf p+q=k}}d{\bf q}\left( \frac{E(q)}{4\pi q^2}\right) \times  \\ \left[ 
\frac{a_3(k,p,q)B_0^{+-}(p)-a_1(k,p,q)\left( X_0^{\pm \pm }(p)+B_0^{\pm \pm
}(p)\right) }{2X_0^{\pm \pm }(p)\left( kB_0^{\pm \pm }(k)+pX_0^{\pm \pm
}(p)-qX_0^{\pm \pm }(q)\right) }\right] 
\end{array}
\end{equation}
where $2k^2a_i(k,p,q)=A_i(k,p,q)$ and $X_0^{\pm \pm }(k)=\sqrt{(B_0^{\pm \pm
}(k))^2-(B_0^{\pm \mp 2}(k))^2}$. The terms $A_i(k,p,q)$ are given in the
Appendix of Leslie \cite{Lesl} as $B_i(k,p,q)$. Since, $E^{+}=E^{-}$ and $%
r_A=1$, it is clear that $\delta B_0^{++}(k)=\delta B_0^{--}(k)$. Therefore, 
$B_0^{++}(k)=B_0^{--}(k)=B_0(k)$ and $X_0^{++}(k)=X_0^{--}(k)=X_0(k).$

Let us denote $B_1(k)$ as the effective mean magnetic field after the
elimination of the first shell. 
\begin{equation}
B_1(k)=B_0(k)+\delta B_0(k)
\end{equation}
Similarly, 
\begin{equation}
B_1^{+-}(k)=B_0^{+-}(k)+\delta B_0^{+-}(k)
\end{equation}

\item  We keep eliminating the shells one after the other by the above
procedure. After $n+1$ iterations we obtain 
\begin{equation}
\label{Bn}B_{n+1}^{ab}(k)=B_n^{ab}(k)+\delta B_n^{ab}(k)
\end{equation}
where the equations for $\delta B_n^{\pm \pm }(k)$ and $\delta B_n^{\pm \mp
}(k)$ are the same as the equations (\ref{pp},\ref{pm}) except that the
terms $B_0^{ab}(k)$ and $X_0^{ab}(k)$ are to be replaced by $B_n^{ab}(k)$ and%
$\,X_n^{ab}(k)$ respectively$.$ Clearly $B_{n+1}(k)$ is the effective mean
magnetic field after the elimination of the $(n+1)$th shell.

The set of RG equations to be solved are (\ref{pp},\ref{pm}) with $B_0$
replaced by $B_n$s, and (\ref{Bn}).
\end{enumerate}

\subsection{Solution of RG equations}

To solve the Eqs. (\ref{pp},\ref{pm}) with $B_n$s and (\ref{Bn}), we
substitute the following forms for $E(k)$ and $B_n(k)$ in the modified
equations (\ref{pp},\ref{pm}) 
\begin{equation}
\begin{array}{c}
E(k)=K\Pi ^{2/3}k^{-5/3} \\ 
B_n^{ab}(k_nk^{\prime })=K^{1/2}\Pi ^{1/3}k_n^{-1/3}B_n^{*ab}(k^{\prime }) 
\end{array}
. 
\end{equation}
with $k=k_{n+1}k^{\prime }$ ( $k^{\prime }>1$). We expect that $%
B_n^{*ab}(k^{\prime })$ is an universal function for large $n$. We use $\Pi
^{+}=\Pi ^{-}=\Pi $ due to symmetry$.$ After the substitution we obtain the
equations for $B_n^{*ab}(k^{\prime })$ that are 
\begin{equation}
\label{rg1}
\begin{array}{c}
\delta B_n^{*}(k^{\prime })=-\int_{
{\bf p}^{\prime }{\bf +q}^{\prime }{\bf =k}^{\prime }}d{\bf q}^{\prime
}\left( \frac{E(q^{\prime })}{4\pi q^{\prime 2}}\right) \times \\ \left[ 
\frac{a_2(k,p,q)\left( X_n(sp)+B_n(sp)\right) -a_4(k,p,q)B_n^{+-}(sp)}{%
2X_n(sp^{\prime })\left( k^{\prime }B_n(sk^{\prime })+p^{\prime
}X_n(sp^{\prime })-q^{\prime }X_n(sq^{\prime })\right) }\right] 
\end{array}
\end{equation}
\begin{equation}
\label{rg2}
\begin{array}{c}
\delta B_n^{*+-}(k^{\prime })=-\int_{
{\bf p}^{\prime }{\bf +q}^{\prime }{\bf =k}^{\prime }}d{\bf q}^{\prime
}\left( \frac{E(q^{\prime })}{4\pi q^{\prime 2}}\right) \times \\ \left[ 
\frac{a_3(k,p,q)B_n^{+-}(sp^{\prime })-a_1(k,p,q)\left( X_n(sp^{\prime
})+B_n(sp^{\prime })\right) }{2X_n(sp^{\prime })\left( k^{\prime
}B_n(sk^{\prime })+p^{\prime }X_n(sp^{\prime })-q^{\prime }X_n(sq^{\prime
})\right) }\right] 
\end{array}
\end{equation}
\begin{equation}
\label{rg3}B_{n+1}^{*ab}(k)=s^{1/3}B_n^{*ab}(k)+s^{-1/3}\delta B_n^{*ab}(k) 
\end{equation}
Now we need to solve these three equations self consistently. The integrals
in the Eqs. (\ref{rg1},\ref{rg2}) is performed over a region $1/s\leq
p^{\prime },q^{\prime }\leq 1$ with the constraint that ${\bf p}^{\prime }%
{\bf +q}^{\prime }{\bf =k}^{\prime }$. We use Monte Carlo technique to solve
the integral. Since the integrals are identically zero for $k^{\prime }>2$,
the initial $B_0^{*}(k_i^{\prime })=B_0^{initial}$ for $k_i^{\prime }<2$ and 
$B_0^{*}(k_i^{\prime })=B_0^{initial}*(k_i^{\prime }/2)^{-1/3}$ for $%
k^{\prime }>2$. We take $B_0^{+-}=0$. The Eqs. (\ref{rg1},\ref{rg2}) are
solved iteratively. We continue iterating the equations till $%
B_{n+1}^{*}(k^{\prime })\approx B_n^{*}(k^{\prime })$, that is, till the
solution converges. For $B_0^{initial}=1.0,$ the $B_n^{\prime }$s for
various $n$ ranging from $0 \ldots 3$ is shown in Figure 1. Here the
convergence is very fast, and after $n=3-4$ iterations $B_n^{*}(k)$
converges to an universal function 
$$
f(k^{\prime })=1.24*k^{\prime -0.32}. 
$$
From the above arguments, we have shown that $B_n^{*}(k^{\prime })\ $is
approximately proportional to $k^{\prime -1/3}$. The other parameter $%
B_n^{*+-}(k^{\prime })$ remains close to zero.

We infer from the above analysis that the mean magnetic field scales as $%
k^{-1/3}$, and the energy spectra scales as $k^{-5/3}$. Essentially, the
scaling of $B_0$ leads to $k^{-5/3}$ energy (Kolmogorov-like) spectra in our
scheme. We have calculated $B_n^{*}(k^{\prime })$ for $B_0^{initial}=1,2,10$
and found that for large $n$, $B_n^{*}(k^{\prime })\approx
1.25B_0^{initial}k^{\prime -1/3}$ or 
\begin{equation}
B_n(k)=1.25B_0^{initial}K^{1/2}\Pi ^{1/3}k^{-1/3}. 
\end{equation}

\subsection{Calculation of $K$}

We can calculate the Kolmogorov's constant for MHD turbulence $K$ by
calculating the cascade rate $\Pi $ \cite{Lesl}. In MHD the cascade rates
are 
\begin{equation}
\Pi ^{+}(k)=\Pi ^{-}(k)=-\int_0^kdk^{\prime }T(k^{\prime }) 
\end{equation}
The numerical solution of the cascade rate integral yields \cite{Lesl} 
\begin{equation}
\label{alpha}\frac{1.24B_0^{initial}}{K^{3/2}}=3.85 
\end{equation}
From the above equation it is evident that the Kolmogorov's constant $K$ is
dependent on the mean magnetic field $B_0^{initial}$, in fact, $K\propto
(B_0^{initial})^{2/3}$. Clearly, an increase in the mean magnetic field
leads to an increase in the Kolmogorov constant, which in turn will lead to
a decrease in the cascade rate (cf. Eq. (\ref{eq:mhd})). This result is
consistent with the simulation results of Oughton \cite{Ough}. However, a
cautious remark is necessary here. We have considered the mean magnetic
field to be isotropic; this isotropy assumption needs to be relaxed for
studies of realistic situations.

\section{Conclusions}

We obtain Kolmogorov-like energy spectrum in MHD turbulence in presence of
arbitrary $B_0$ by postulating that the effective $B_0$ is scale dependent
In our renormalization group scheme we find that the self consistent $B_n(k)$
is proportional to $k^{-1/3}$ and $E(k)$ is proportional to $k^{-5/3}$. This
analysis has been worked out when $E^{+}=E^{-}$ and $r_A=1.$ The
generalization to arbitrary parameters is planned for future studies.

In our methodology, the averaging has been performed for small wavenumbers
in contrast to earlier RG analysis of turbulence in which the higher
wavenumbers were averaged out. Our scheme yields a power law solution for
large wavenumber, and is independent of the small wavenumber forcing states.
This is in agreement with the Kolmogorov's hypothesis which states that the
energy spectrum of the intermediate scale is independent of the large-scale
forcing. Any extension of our scheme to fluid turbulence in presence of
large-scale shear etc. will yield interesting insights into the connection
of energy spectrum with large-scale forcing.

I thank V. Subrahmanyam, J. K. Bhattacharjee, and M. Barma for numerous
useful discussions.

\nopagebreak
\begin{figure}
\figure{Figure 1: $B_n^{*}(k^{\prime })$ for $n=0 \ldots 3$. The line of best fit $f(k^{\prime })$ to $B_3^{*}(k^{\prime })$ overlaps with $B_3^{*}$.}
\end{figure}

\end{document}